\documentclass[sigconf]{acmart}

\usepackage{graphicx}
\usepackage{listings}
\usepackage{xcolor}
\usepackage{graphicx}
\usepackage{amsmath}

\usepackage{enumitem}

\definecolor{salmon}{RGB}{250,128,114}
\newcommand{\jz}[1]{\textcolor{black}{#1}}

\definecolor{softblue}{RGB}{140, 178, 210}

\setcopyright{none}
\acmConference[DAC '26]{63rd ACM/IEEE Design Automation Conference}{July 26--29, 2026}{Long Beach, CA, USA}

\renewcommand\footnotetextcopyrightpermission[1]{}

\begin{document}

\title{FILCO: \underline{F}lex\underline{i}b\underline{l}e \underline{Co}mposing Architecture with Real-Time Reconfigurability for DNN Acceleration}

\settopmatter{authorsperrow=4}

\author{Xingzhen Chen}
\affiliation{%
  \institution{Brown University} 
  \city{Providence}
  \country{USA}
  }
\email{xingzhen_chen@brown.edu}

\author{Jinming Zhuang}
\affiliation{%
  \institution{Brown University} 
  \city{Providence}
  \country{USA}
  }
\email{jinming_zhuang@brown.edu}

\author{Zhuoping Yang}
\affiliation{%
  \institution{Brown University} 
  \city{Providence}
  \country{USA}
  }
\email{zhuoping_yang@brown.edu}

\author{Shixin Ji}
\affiliation{%
  \institution{Brown University} 
  \city{Providence}
  \country{USA}
  }
\email{shixin_ji@brown.edu}

\author{Sarah Schultz}
\affiliation{%
  \institution{Brown University} 
  \city{Providence}
  \country{USA}
  }
\email{sarah_schultz2@brown.edu}

\author{Zheng Dong}
\affiliation{%
  \institution{Wayne State University}
  \city{Detroit}
  \country{USA}
  }
\email{dong@wayne.edu}

\author{Weisong Shi}
\affiliation{%
  \institution{University of Delaware}  
  \city{Newark}
  \country{USA}
  }
\email{weisong@udel.edu}

\author{Peipei Zhou}
\affiliation{%
  \institution{Brown University} 
  \city{Providence}
  \country{USA}
  }
\email{peipei_zhou@brown.edu}

\renewcommand{\shortauthors}{
Xingzhen Chen,
Jinming Zhuang,
Zhuoping Yang,
Shixin Ji,\\
Sarah Schultz,
Zheng Dong,
Weisong Shi,
and Peipei Zhou
}

\begin{abstract}
With the development of deep neural network (DNN) enabled applications, achieving high hardware resource efficiency on diverse workloads is non-trivial in heterogeneous computing platforms. 
Prior works discuss dedicated architectures to achieve maximal resource efficiency. However, a mismatch between hardware and workloads always exists in various diverse workloads. 
Other works discuss overlay architecture that can dynamically switch dataflow for different workloads. 
However, these works are still limited by flexibility granularity and induce much resource inefficiency.

To solve this problem, we propose a flexible composing architecture, FILCO, that can efficiently match diverse workloads to achieve the optimal storage and computation resource efficiency. FILCO can be reconfigured in real-time and flexibly composed into a unified or multiple independent accelerators. We also propose the FILCO framework, including an analytical model with a two-stage DSE that can achieve the optimal design point.
We also evaluate the FILCO framework on the 7nm AMD Versal VCK190 board. Compared with prior works, our design can achieve $1.3\times$$\sim$$5\times$ throughput and hardware efficiency on various diverse workloads.

\vspace{-5pt}

\end{abstract}

\maketitle

\begingroup
\renewcommand\thefootnote{}
\footnotetext{This paper has been accepted to the 63rd ACM/IEEE Design Automation Conference (DAC).}
\endgroup

\section{Introduction}

In many real-world applications, an end-to-end task must execute a variety of deep neural networks (DNNs) with fundamentally different characteristics in both computation and communication patterns. 
However, modern accelerators often adopt specific dataflows tailored for certain DNN types, resulting in significant performance degradation when handling workloads with diverse requirements.
For example, in autonomous driving systems (ADS), there are MLPs for multilayer perceptron classification or regression~\cite{wang2019benchmarking}, DeiT for image segmentation~\cite{DeiT}, MLP-Mixer for image classification~\cite{MLPMixer}, and PointNet for 3-D point-cloud processing~\cite{qi2017pointnet}. 
Dense matrix multiply (MM) operations appear to be computation-intensive, but small matrix dimensions can shift the bottleneck to communication. 
Their dimensions also vary dramatically from layer to layer, resulting in large intra-model shape variance.
Moreover, since different sub-tasks require varying accuracy levels, DNNs of different types and sizes are deployed, resulting in inter-model shape variance.

\begin{figure}
    \centering
    \includegraphics[width=1\linewidth]{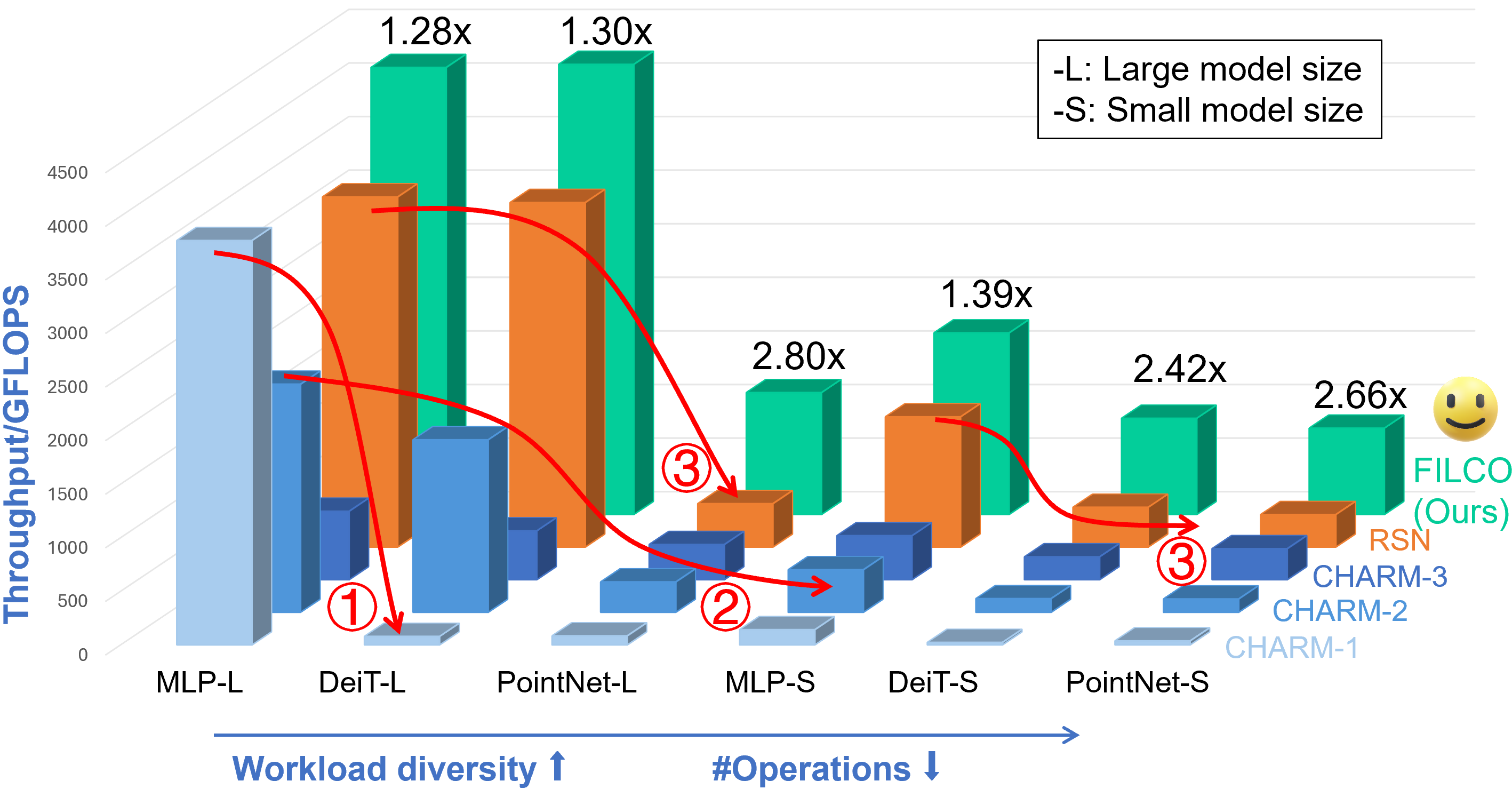}
    \vspace{-20pt}
    \caption{Throughput comparison for different works.}
    \label{fig: motivation figure}
    \vspace{-20pt}
\end{figure}

As shown in Figure \ref{fig: motivation figure}, we profile several existing accelerators for DNN models, and we choose three models with varying degrees of diversity. 
In the example, we select MLP~\cite{wang2019benchmarking} to represent a DNN with low intra-model diversity, as it primarily consists of matrix multiplication layers with near-square shapes.
In contrast, DeiT~\cite{DeiT} is a Transformer-based model with a medium degree of diversity due to the different shapes of its attention and feed-forward layers. PointNet~\cite{qi2017pointnet} exhibits the highest diversity because of its T-Net architecture.
Workload diversity also exists across models of the same type but with different sizes. For example, MLP-L and MLP-S share a similar architecture yet exhibit a high degree of diversity due to inter-model variation.
We also profile performance on MLP, DeiT, and PointNet models with varying sizes to showcase the capacity for inter-model diversity.
CHARM-1 is one of the monolithic designs in CHARM~\cite{zhuang2023charm}, which fully utilizes the on-chip resources on the Versal platform~\cite{versal_acap}. According to profiling results, CHARM-1 can achieve high throughput for the MLP-L model due to efficient resource utilization for large and non-diverse MM. However, when the workload switches to the model with a higher degree of diversity (DeiT-L) or with a smaller size (MLP-S), the throughput degrades rapidly (\textcircled{1}). Although CHARM-1 fully utilizes resources, it has to pad operand matrices to the fixed on-chip buffer size when executing smaller and diverse workloads, inducing much communication and computation overhead. Designing two diverse accelerators to suit diverse operand shapes can be a solution, as shown in CHARM-2 and CHARM-3. From the profiling results, such a design methodology can achieve steady performance degradation for smaller model sizes and higher degrees of diversity (\textcircled{2}). However, since they reallocate a small portion of resources for a small accelerator, CHARM-2 and CHARM-3 cannot achieve as high throughput as CHARM-1 can in the large and non-diverse models (MLP-L). 
Since the resource partition must be determined before runtime and lacks reconfigurability, the trade-off always happens in a series of static accelerator designs.

Existing work~\cite{hpca21herald} attempts to address this issue by instantiating multiple sub-accelerators for dedicated workloads, but they only support limited dataflow and lack customization for different workloads.
Other existing works have tried to solve it in an overlay fashion, e.g., RSN~\cite{RSN}. Since RSN can flexibly map operand matrices to on-chip buffers and concatenate computation tiles for diverse workloads, it can alleviate this problem to some extent. However, RSN is limited by its static on-chip matrix shape and fixed computation tile size across computation cores. As a result, it can sustain high efficiency only when the model size is large and the degree of diversity remains relatively low. As shown in Figure \ref{fig: motivation figure} \textcircled{3}, RSN can sustain better throughput from MLP-L to DeiT-L than CHARM, but when the model size becomes smaller and the workload diversity increases, RSN suffers from a sharp drop in performance.

In order to unlock the full flexibility for hugely diverse workloads, it is essential and non-trivial to explore the flexibility and reconfigurability on both the computation side and the communication side. On the computation side, this requires designing computation logic with runtime-flexible computing tile sizes to eliminate additional overhead when launching each computation core in diverse workloads. On the communication side, in order to avoid unnecessary off-chip access, it requires designing a flexible on-chip memory that can be configured to store diverse operand shapes without extra padding overhead. In addition, flexible mapping between operand matrices and on-chip memory is essential to improve the alignment between workloads and the underlying hardware. To address these challenges, we propose \textbf{FILCO}, a flexible composing architecture with real-time reconfigurability for DNN acceleration. We summarize our contributions as follows: 

\vspace{-2.5pt}
\begin{itemize}[leftmargin=*]
    \item On the hardware side, we propose an AIE programming method that supports fine-grained, runtime-flexible parallelism patterns and an on-chip memory management method to avoid communication overhead.
    \item On the algorithm side, we formulate DSE into a two-stage optimization flow and apply MILP to search for the optimal design point, as well as a GA heuristic to reduce DSE search time.
    \item We propose the FILCO framework that can take DNN models, platform information, and DDR profiling results as input, and generate the ready-to-run binary files as output. We deploy our framework and conduct experiments on diverse workloads, which can achieve $1.3\times$$\sim$$5\times$ gains compared with existing works.
\end{itemize}

\vspace{-5pt}
\section{FILCO Architecture}

In this section, we first introduce \jz{the overview of FILCO architecture in Section \ref{sec: FILCO Architecture Overview}. Then, we discuss the three proposed hardware methodologies} in Sections \ref{sec: Flexible Computation Parallelism}, \ref{sec: Flexible On-chip Memory View}, and \ref{sec: Flexible On-chip Memory Functionality}. In Section \ref{sec: Instruction}, we explain the control flow and instruction set in FILCO.

\vspace{-5pt}
\subsection{Architecture Overview}
\label{sec: FILCO Architecture Overview}

\begin{figure}
    \centering
    \includegraphics[width=0.8\linewidth]{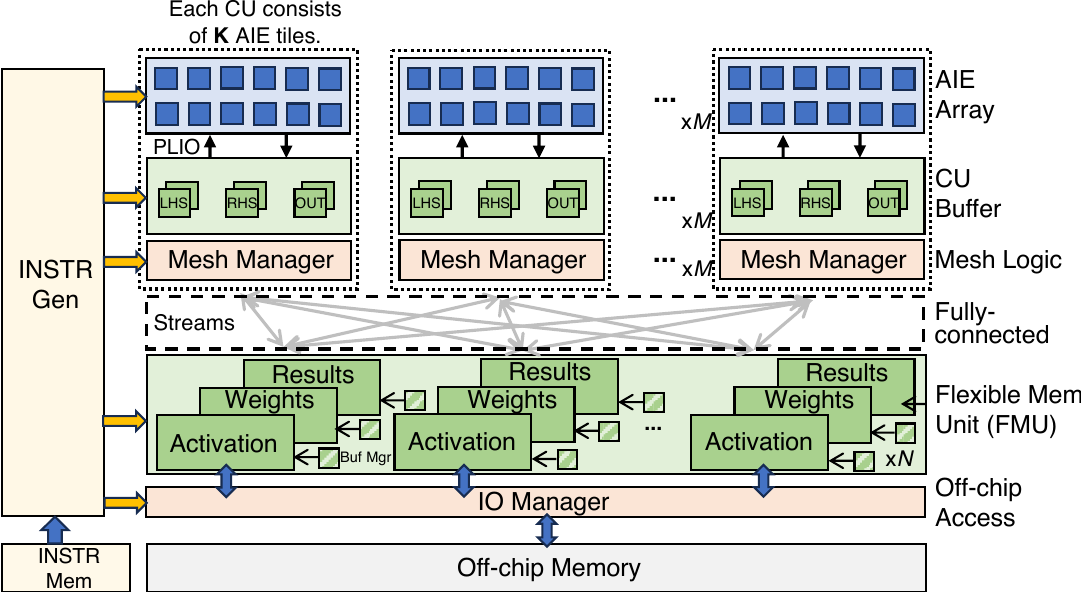}
    \vspace{-5pt}
    \caption{FILCO hardware architecture.}
    \vspace{-15pt}
    \label{fig:architecture}
    
\end{figure}

The FILCO hardware architecture overview is shown in Figure \ref{fig:architecture}. 
\jz{It is mainly composed of a data plane, responsible for computation and data movement, and a control plane, which manages instruction generation and execution scheduling.} In the data plane, we partition the on-chip resources into Compute Units (CU), Flexible Memory Units (FMU), and IO Manager (IOM). \jz{Each Compute Unit is featured with an AI Engine (AIE) array, a CU Buffer, and a Mesh Manager, and is responsible for handling the compute-intensive workloads.} 
\jz{The Flexible Memory Units explore data reuse by allocating on-chip buffers on the Programmable Logic (PL)}. Additionally, the IO Manager is to handle the data communication between FMUs and off-chip memory.
\jz{The execution of each unit in the data plane is controlled by the instruction sets proposed by FILCO}.
In the control plane, the Instruction Generator loads instructions from \jz{off-chip Instruction Memory and dispatches them} to each function unit according to the instruction header. \jz{Each function unit first receives and decodes the instruction, then executes it according to the control signals in the instructions.}

\jz{In our proposed FILCO architecture}, there are \textit{N} FMUs and \textit{M} CUs. Each CU consists of \textit{K} AIEs. 
\jz{The function units are connected through multi-level streaming hierarchies. For the off-chip communication, we design IO Managers that enable different FMUs to access a unified memory space. For the on-chip communication, we apply a fully-connected stream topology between the FMUs and CUs to achieve maximum flexibility that can adapt to diverse workloads. Within each CU, there is a Mesh Manager to handle the mesh-in and mesh-out logic control.}


To sustain both off-chip and on-chip bandwidth, proper buffer partitioning is required to avoid bank conflicts.
We adopt wide port widths for off-chip access with cyclic partitioning, while using block partitioning to match the tiled MM execution of AIE and prevent on-chip buffer conflicts.
To decouple partitioning conflicts between the AIE and IO Manager, we introduce a hierarchical on-chip memory, including CU Buffer and FMU.
In FILCO, CU Buffers are sized to match the maximum AIE tile and use block partitioning, while FMU size depends on available on-chip buffers and adopts cyclic partitioning for wide ports.
Next, we detail the hardware design methodologies in the following sections.


\vspace{-5pt}
\subsection{Flexible Computation Parallelism}
\label{sec: Flexible Computation Parallelism}

In FILCO design, we \jz{assign} the computation-intensive workloads to AIE. In order to achieve high efficiency on diverse workloads without reloading bitstreams at runtime, we propose a flexible AIE programming method that can switch between different execution modes at runtime to achieve flexible parallelism patterns.

\begin{figure}[t]
\centering

\begingroup
\lstset{
    basicstyle=\ttfamily\scriptsize\color{black},
    commentstyle=\color[rgb]{0.0,0.5,0.0},
    stringstyle=\color[rgb]{0.6,0.2,0.0},
    breaklines=true,
    columns=fullflexible,
    showstringspaces=false,
    emph={bound_i,bound_k,bound_j},
    emphstyle={\color{red}},
    numbers=left,
    numberstyle=\tiny, 
    numbersep=1pt, 
}

\begin{minipage}[t]{0.48\linewidth}

\begin{lstlisting}[language=C++]
void singleAIE_kernel(input_buffer& in0, input_buffer& in1, output_buffer& out0) {
    // Load instr vec from local buf
    vector<3> Instr = load_v<3>(in0);
    // Assign flexible loop boundary
    int bound_i = Instr.get(0);
    int bound_k = Instr.get(1);
    int bound_j = Instr.get(2);
    // Main loop
    for(int i=0; i < bound_i; i++){
      for(int j=0; j < bound_j; j++){
        for(int k=0; k < bound_k; k++){
          //Atomic operation for 2x8x8
          ...}}}}
\end{lstlisting}
\centering
(a)AIE kernel pseudo-code
\end{minipage}
\hspace{0pt}
\begin{minipage}[t]{0.48\linewidth}
\vspace{0pt}
\centering
\includegraphics[width=0.7\linewidth]{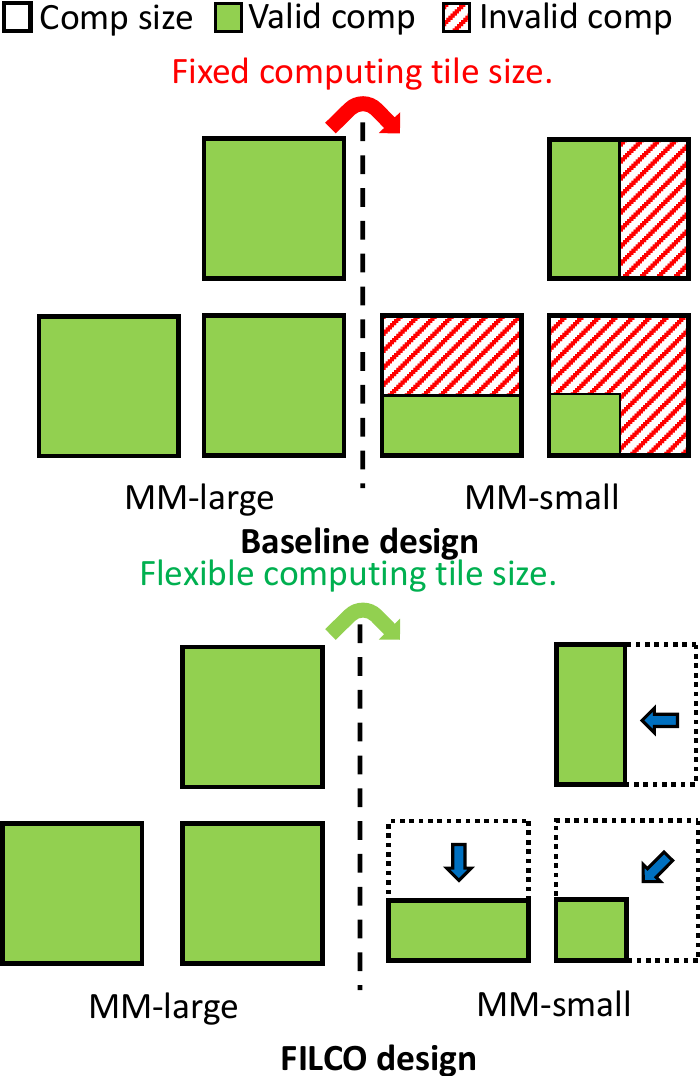}
\centering
(b)Flexible computing tile sizes
\end{minipage}
\endgroup
\vspace{0pt}
\vspace{-10pt}
\caption{Flexible parallelism and single AIE programming.}
\label{fig: flexible parallelism}
\vspace{-10pt}
\end{figure}

Existing works either use static AIE instructions for peak efficiency or finite instruction blocks for mode switching.
However, static instructions lead to performance degradation on diverse workloads.
As shown in Figure \ref{fig: flexible parallelism}(b), the compute tile size remains fixed across iterations. While large workloads fully utilize parallelism (green blocks), smaller workloads require padding, resulting in significant invalid computation (red blocks).
Designing finite instruction blocks helps to mitigate the invalid computation, but it has significant limitations in practice. There are only 16KB of instruction memory in each AIE, and the instruction size for computing MM with a tile size of 32x32x32 is more than 4KB.


In FILCO, to improve flexibility and efficiency, we pack each 2×8×8 tiled matrix multiplication into an atomic operation to maintain high VLIW efficiency (Line 13 in Figure \ref{fig: flexible parallelism}), and define nested \texttt{for} loops with dynamic boundaries to enable flexible tile sizes (Lines $10$$\sim$$12$). The loop boundaries are provided through input ports (Lines $3$$\sim$$7$).
At runtime, tile sizes are configured by issuing different instructions, as shown in Figure \ref{fig: flexible parallelism}(b). For large workloads, the outer loop boundaries are adjusted to fully utilize compute resources. For small workloads, reconfiguration reduces tile sizes to match workloads, significantly minimizing unnecessary computation (green blocks in FILCO).


\begin{figure}[t]
\centering

\begingroup
\lstset{
    basicstyle=\ttfamily\scriptsize\color{black},
    commentstyle=\color[rgb]{0.0,0.5,0.0},
    stringstyle=\color[rgb]{0.6,0.2,0.0},
    breaklines=true,
    columns=fullflexible,
    showstringspaces=false,
    emph={count,startRow,endRow,startCol,endCol,ping},
    emphstyle={\color{red}},
    numbers=left,
    numberstyle=\tiny, 
    numbersep=1pt, 
}

\begin{minipage}[t]{0.48\linewidth}

\begin{lstlisting}[language=C++]
void FMU_kernel(stream &Instr, 
stream &IOM2FMU,stream &FMU2CU0,..)
{ //Instantiate 1-D double buffer
  buf0[buf_size]; buf1[buf_size];
  //Cyclic partition
  #pragma var=buf0 buf1 type=cyclic
  while(1){
    //Load instr in control plane 
    instr = Instr.read();
    if(instr.ping){
    //Diff buf view based on instr
    buf0_recFromIOM(instr.count,..);
    buf1_sendToCU(instr.startRow,instr.endRow,instr.startCol,instr.endCol,..);
    }else{.../*pong operations*/}}}

\end{lstlisting}
\centering
(a)FMU pseudo-code
\end{minipage}
\hspace{0pt}
\begin{minipage}[t]{0.48\linewidth}
\vspace{0pt}
\centering
\includegraphics[width=0.7\linewidth]{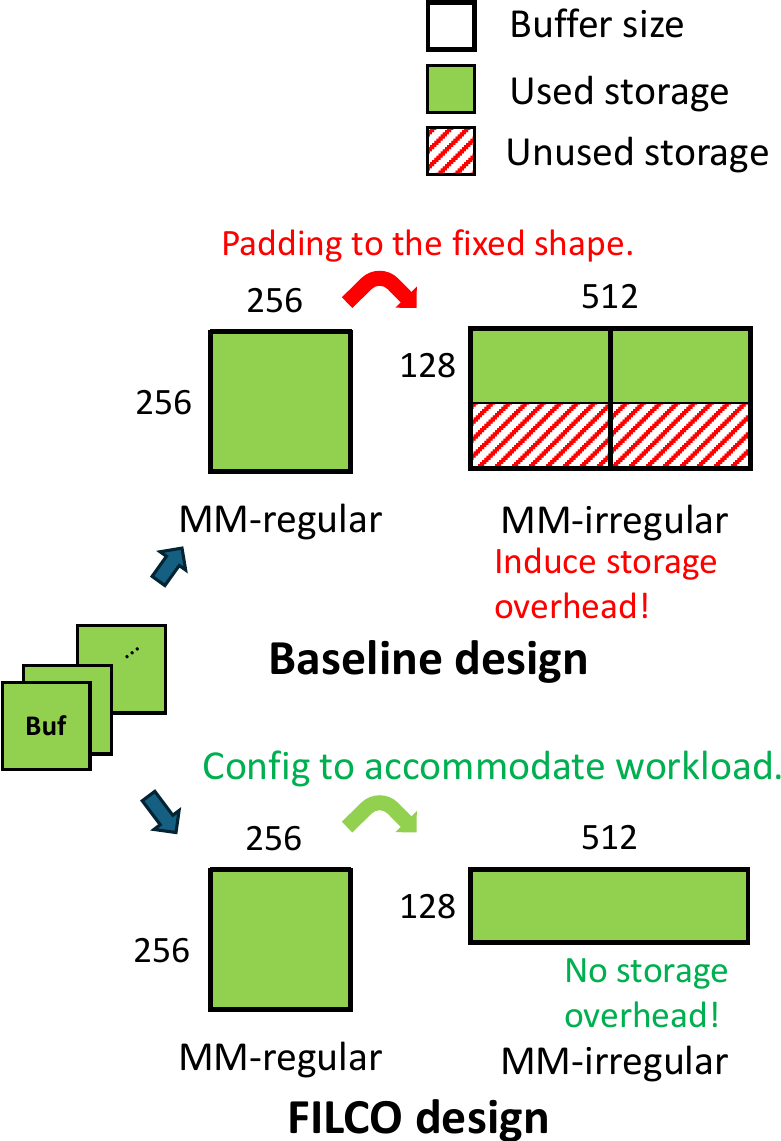}
\centering
(b)Flexible memory views
\end{minipage}
\endgroup
\vspace{0pt}
\vspace{-10pt}
\caption{Flexible on-chip memory views.}
\label{fig: flexible memory view}
\vspace{-15pt}
\end{figure}

\vspace{-5pt}
\subsection{Flexible On-chip Memory View}
\label{sec: Flexible On-chip Memory View}

As shown in Figure \ref{fig: flexible memory view}(b), existing accelerators typically use an N-dimensional on-chip buffer that matches operand dimensions.
However, such static buffer views lead to poor storage utilization for diverse workloads in practice.
For example, as presented in the baseline design in Figure \ref{fig: flexible memory view}(b), assume that we have a 2-D matrix with size 256x256 to be stored on-chip. In existing static accelerator designs, they instantiate a static buffer shape of 256x256 to perfectly match the workload shape, which can achieve a high efficiency in certain static workloads (green block). However, when they are required to handle diverse workloads, e.g., 128x512 matrix shapes, such a static design method induces much storage overhead, and only achieves 50\% efficiency due to unnecessary padding (red block). In reality, the two diverse matrices have the same data size, which can definitely be stored in one buffer.

Therefore, proposing a flexible on-chip memory that is able to switch between different buffer views can improve efficiency. As shown in Figure\ref{fig: flexible memory view}(a), FILCO instantiates multiple parallel FMUs containing a 1-D addressing double-buffer and an instruction decoder (Line 5 in Figure \ref{fig: flexible memory view}). When the system is launched, each FMU instance starts to execute simultaneously and decode instructions from the control plane (Lines 9-11). In the receive stage, FMU only receives a given number of elements that is determined from instructions (Line 14). In the send stage, FMU decodes the tile information to address it from 1-D indexing and sends it to the correct downstream function units through pre-routed streams (Line 15). 
After enabling such flexible on-chip memory views, FMU can be viewed as diverse shapes to perfectly match workloads, which can reduce much communication overhead and improve storage efficiency.

\vspace{-5pt}
\subsection{Flexible On-chip Memory Functionality}
\label{sec: Flexible On-chip Memory Functionality}

\begin{figure}
    \centering
    \includegraphics[width=0.6\linewidth]{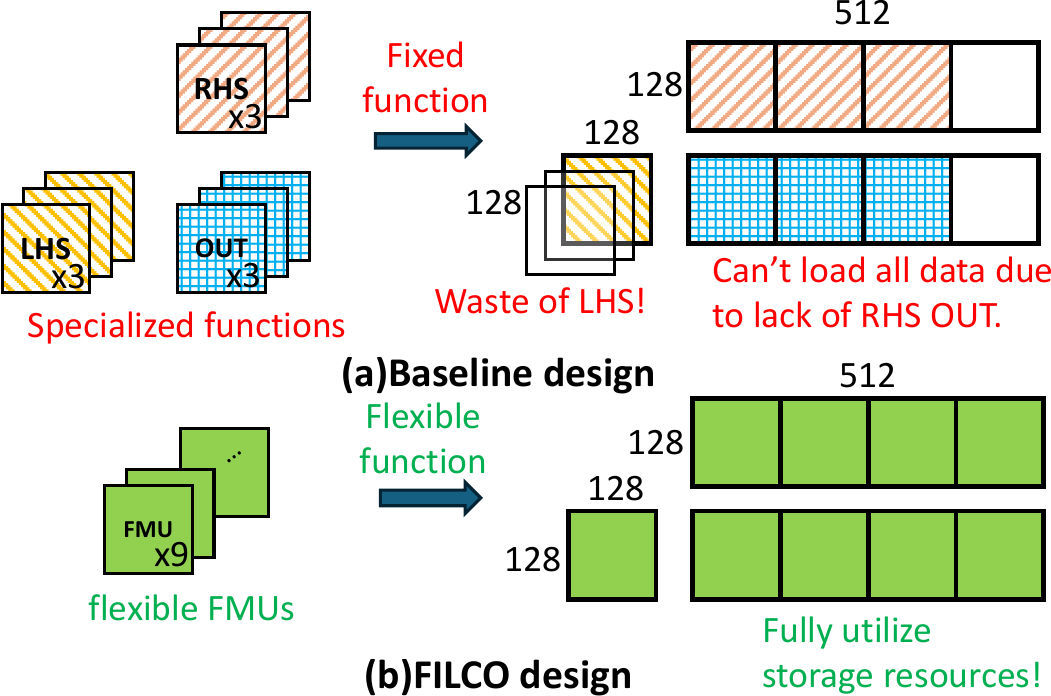}
    \vspace{-10pt}
    \caption{Flexible on-chip memory functionality.}
    \vspace{-18pt}
    \label{fig: flexible memory func}

\end{figure}

To maximize on-chip memory reuse, FILCO introduces FMUs with flexible memory allocation.
Unlike prior designs that statically allocate weight and activation buffers at compile time, FMUs can be dynamically configured at runtime to support diverse workloads.
In practice, workloads may have one dimension much larger than others in MM. Static buffer allocation cannot handle this efficiently, as larger dimensions require more storage for certain operands, exceeding buffer limits and preventing full data loading (Figure \ref{fig: flexible memory func} (a)).
In FILCO, to avoid the limitation of \jz{specifying buffers for a certain operand}, our proposed FMU is able to be configured with different functionalities for operands or results based on control instructions. As shown in Figure \ref{fig:architecture}, we connect each FMU instance with the IO Manager and CUs using pre-routed streams to construct the flexible dataflow plane. Each FMU has an independent instruction decoder to configure according to the instructions.
For diverse workloads, FILCO can maximize data reuse as long as the total data size of operands and results matrices does not exceed resource constraints in Figure \ref{fig: flexible memory func} (b).

\vspace{-5pt}
\subsection{Instruction Set}
\label{sec: Instruction}

In this section, we elaborate on FILCO control flow and instruction set design. 
FILCO distinguishes between static and runtime parameters. Static parameters are fixed before compilation, such as the number and capacity of FMUs/CUs and AIE connections within a CU.
Runtime parameters can be configured at execution time through instruction decoding. For example, FMUs can adapt to different data sizes, with patterns switched by decoding a few bytes of instructions. Therefore, an instruction set enabling runtime configurability is essential for supporting diverse workloads.

Table \ref{tab: instruction set} lists the instruction sets for different function units. In general, Instruction Generator loads the instruction header from off-chip memory, which contains the valid instruction length in the instruction sequence and the destination units. \jz{Based on the DDR address from the instructions, the IO Managers achieve high DDR bandwidth by issuing AXI transactions with large burst length.}
\jz{The FMUs use the correct src/des units and 1-D addressing control information provided by the instructions to gather and scatter data, thereby maximizing the on-chip data reuse.}
CUs perform computation operations and are responsible for loading operands from correct FMUs and storing results to correct FMUs, as well as achieving high computation resource efficiency.

\begin{table}[t]

\caption{Instruction sets for function units.}
\vspace{-10pt}
\footnotesize
\centering
\resizebox{\columnwidth}{!}{%
\begin{tabular}{ll}
\toprule[1.2pt]
\textbf{Function Unit}     & \textbf{Instruction Items} \\
\midrule
Instr Generator   & is\_last, des\_unit, valid\_length \\
\midrule
IOM Loader &
\begin{tabular}[t]{@{}l@{}}
is\_last, ddr\_addr, des\_fmu, M, N,\\
start\_row, end\_row, start\_col, end\_col
\end{tabular}
\\
\midrule
IOM Storer &
\begin{tabular}[t]{@{}l@{}}
is\_last, ddr\_addr, src\_fmu, M, N,\\
start\_row, end\_row, start\_col, end\_col
\end{tabular}
\\
\midrule
FMU &
\begin{tabular}[t]{@{}l@{}}
is\_last, ping\_op, pong\_op, src\_cu, des\_cu, count,\\
start\_row, end\_row, start\_col, end\_col
\end{tabular}
\\
\midrule
CU & is\_last, ping\_op, pong\_op, src\_fmu, des\_fmu, count \\
\bottomrule[1.2pt]
\end{tabular}%
}

\label{tab: instruction set}
\vspace{-15pt}
\end{table}

\vspace{-5pt}
\section{Analytical Model}

In this section, \jz{we first introduce the FILCO framework overview and the proposed} two-stage design space exploration (DSE) in Section \ref{sec: Two-stage Design Space Exploration}. \jz{We formulate the DSE problem} into an MILP formulation in Section \ref{sec: ILP Formulation} to guarantee the optimality. \jz{To reduce search time, we propose a Genetic Algorithm–based heuristic in Section \ref{sec: GA}.}

\vspace{-5pt}
\subsection{Two-stage Design Space Exploration}
\label{sec: Two-stage Design Space Exploration}

\begin{figure}
    \centering
    \includegraphics[width=0.75\linewidth]{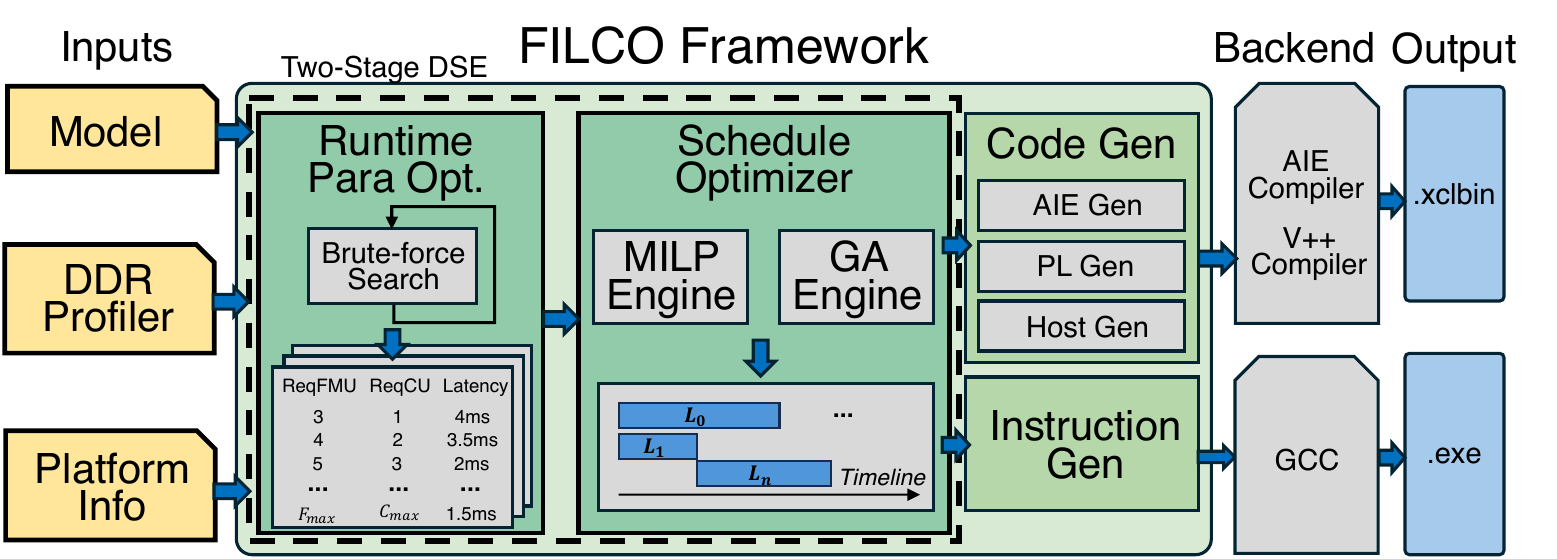}
    \vspace{-10pt}
    \caption{FILCO framework overview.}
    \label{fig: framework}
    \vspace{-15pt}
\end{figure}

Figure \ref{fig: framework} shows an overview of the FILCO framework. FILCO takes DNN models, platform information, and DDR profiling results as input. 
After the automated optimization flow and code generation, FILCO generates the binary files by launching the backend compilers. In the first stage, Runtime Parameter Optimizer performs a brute-force search on every layer to find the optimal runtime dataflow, as well as a table with the optimal latency under the constraints of FMU and CU.
In the second stage, the Schedule Optimizer searches for the optimal schedule timeline, based on the recorded table in the first stage, while guaranteeing that the resource requirements are always \jz{under hardware resource constraints}.
After two-stage DSE optimization, FILCO generates a valid schedule as well as customized dataflow for each DNN layer. Then the Code Generator and Instruction Generator \jz{emit} the HLS codes and instruction sequences for backend compilers based on the scheduling timeline and the recorded runtime dataflow information. 




\vspace{-5pt}
\subsection{MILP Formulation}
\label{sec: ILP Formulation}

\jz{FILCO proposes a two-stage optimization algorithm to solve the coupled mapping and scheduling problem. It ensures optimality by exhaustively enumerating possible layer-to-accelerator candidates, and then formulating an MILP to solve the scheduling stage.}
\jz{After enumerating the possible mapping candiates in Stage 1}, the scheduling optimization flow in FILCO can be formulated as \jz{the following problem \cite{milp_basic, zhou2019customized}:} Given a Directed Acyclic Graph (DAG) for a workload, each node stands for one layer ($L_i$), and each edge stands for the dependency between two layers ($P_{i,j}=1$, if $L_j$ depends on $L_i$). For i-th layer, there are multiple candidate execution modes, and for k-th mode, Runtime Parameter Optimizer records the required number of FMU ($f_{i,k}$), the required number of CU ($c_{i,k}$), the optimal runtime parameters and the optimal latency ($e_{i,k}$). For platform \jz{resource} constraints, there are a total of $F_{max}$ FMUs and $C_{max}$ CUs. 
\jz{Our MILP formulation aims to minimize the overall execution time while preserving application dependencies and ensuring that the hardware utilization remains within the resource constraints.}

We define the decision variables as follows: $A_{i,m}$ and $B_{i,m}$ are binary variables, which are equal to $1$ if $L_i$ uses the m-th FMU or CU, respectively; $M_{i,k}$ is also binary variable and is equal to $1$ if $L_i$ executes in k-th mode; $S_i$ and $E_i$ represent the start time and end time for layer $L_i$, respectively. After defining decision variables, we derive the objective function and constraints.

In a feasible schedule solution, each layer $L_i$ only executes in one certain mode, thus, we have
\vspace{-3pt}
\begin{equation}
\footnotesize
\forall i:\ \sum_k M_{i,k} = 1
\label{constraint 1}
\end{equation}
\vspace{-4pt}

\noindent For the two consecutive layers with $P_{i,j}=1$, the schedule timeline should meet data dependency constraints, i.e., the start time of $L_j$ should be later than the end time of $L_i$, and the end time for $L_i$ is the start time plus the latency of the selected execution mode:
\vspace{-3pt}
\begin{equation}
\footnotesize
\begin{aligned}
\forall (i,j) \in DAG,\, &P_{i,j} = 1: S_j \ge E_i \\
E_i = \sum_k &M_{i,k} \times e_{i,k}
\end{aligned}
\label{constraint 2}
\end{equation}
\vspace{-3pt}

\noindent For the two non-consecutive layers with $P_{i,j}=0$, since each FMU or CU can only execute one layer at one time, if two layers are allocated on the same FMU or CU, the execution in timeline should not overlap. To \jz{represent} the overlap condition, we define a binary variable $O_{i,j}$ where

\vspace{-3pt}
{\footnotesize
\[
O_{i,j} =
\begin{cases}
1, & \text{if } S_i - E_j < 0, \\
0, & \text{if } S_i - E_j \ge 0.
\end{cases}
\]
}

\noindent \jz{To linearize the conditional constraints for the MILP formulation, we introduce a large enough integer $\phi$ and rewrite the constraints in linear form.} 

\vspace{-5pt}
\begin{equation}
\footnotesize
\begin{aligned}
S_i - E_j &< \phi\times(1 - O_{i,j})
\qquad
S_i - E_j &\ge -\phi\times O_{i,j}
\end{aligned}
\label{constraint 3}
\end{equation}

\noindent According to the \jz{definition} of $O_{i,j}$, if schedules of $L_i$ and $L_j$ are overlap in timeline, meaning that $S_i < E_j$ and $S_j < E_i$, \jz{then} $O_{i,j} = O_{j,i} = 1$. Therefore, for any pair of layers with $P_{i,j}=0$ that occupies the same FMU or CU, we have

\vspace{-5pt}
\begin{equation}
\footnotesize
\begin{aligned}
&\forall (i,j) \in DAG, \forall m \in FMU, P_{i,j} = 0 : A_{i,m}+A_{j,m}+O_{i,j}+O_{j,i} \leq 3 \\
&\forall (i,j) \in DAG, \forall m \in CU, P_{i,j} = 0 : B_{i,m}+B_{j,m}+O_{i,j}+O_{j,i} \leq 3
\end{aligned}
\label{constraint 4}
\end{equation}

\noindent If $A_{i,m}+A_{j,m}+O_{i,j}+O_{j,i} > 3$, \jz{it implies} $A_{i,m}=1,A_{j,m}=1,O_{i,j}=1,O_{j,i}=1$. According to the definition of $A_{i,m}$ and $O_{i,j}$, for the pair of $L_i$ and $L_j$, both of them are allocated to m-th FMU, and their schedules overlap, which conflicts with resource constraints.

In order to meet the resource requirements of each candidate mode, we should guarantee that the sum of \jz{utilized} FMU or CU is equal to the \jz{allocated} FMU or CU:

\vspace{-10pt}
\begin{equation}
\footnotesize
\begin{aligned}
\forall i:\ &\sum_m A_{i,m} = \sum_k M_{i,k} f_{i,k}
\qquad
\forall i:\ &\sum_m B_{i,m} = \sum_k M_{i,k} c_{i,k}
\end{aligned}
\label{constraint 5}
\end{equation}

\noindent The object is to find the feasible schedule with optimal latency:

\vspace{-10pt}
\begin{equation}
\footnotesize
\begin{aligned}
\min T \quad \forall i:T \geq E_i
\end{aligned}
\label{constraint 6}
\end{equation}

\noindent Combining Equations~\ref{constraint 1}--\ref{constraint 6}, we can formulate the scheduling optimization flow in FILCO DSE into an MILP problem.

\vspace{-5pt}
\subsection{Genetic Algorithm}
\label{sec: GA}

The main challenge in the two-stage optimization flow is the extremely large design space for scheduling. For example, the complexity for \jz{enumerating only the} candidate mode selection is $O(m^n)$, where $m$ is the number of candidate modes and $n$ is the number of layers in the workload DAG. 
To solve this problem, we propose a \jz{Genetic Algorithm-based heuristic search for Scheduling Optimizer in FILCO DSE.}
In our Genetic Algorithm, we first encode our decision variables into one chromosome and initialize the population with encoded chromosomes. Then we set a maximal iteration time, and within each iteration, every chromosome in the population applies crossover and mutation to generate a new generation. It then evaluates the fitness score of the child's chromosome, saves the one with the best fitness value, and iterates for the next generation.

In the FILCO framework, each design point is defined as a chromosome with $2N$ decision variables, where $N$ is equal to the number of layers. The first $N$ variables are defined as Encode[N], which are real numbers between 0 and 1; The second $N$ variables are defined as Candidate[N], which are integers between 0 and (\#Can $-$ 1), where \#Can is the number of possible execution candidates.
In the rossover and mutation stages, we apply a random selection strategy.
In the decode and evaluation stage, we apply a dependency-aware decoding method to make sure that the dependency is resolved.

\begin{figure}
    \centering
    \includegraphics[width=0.6\linewidth]{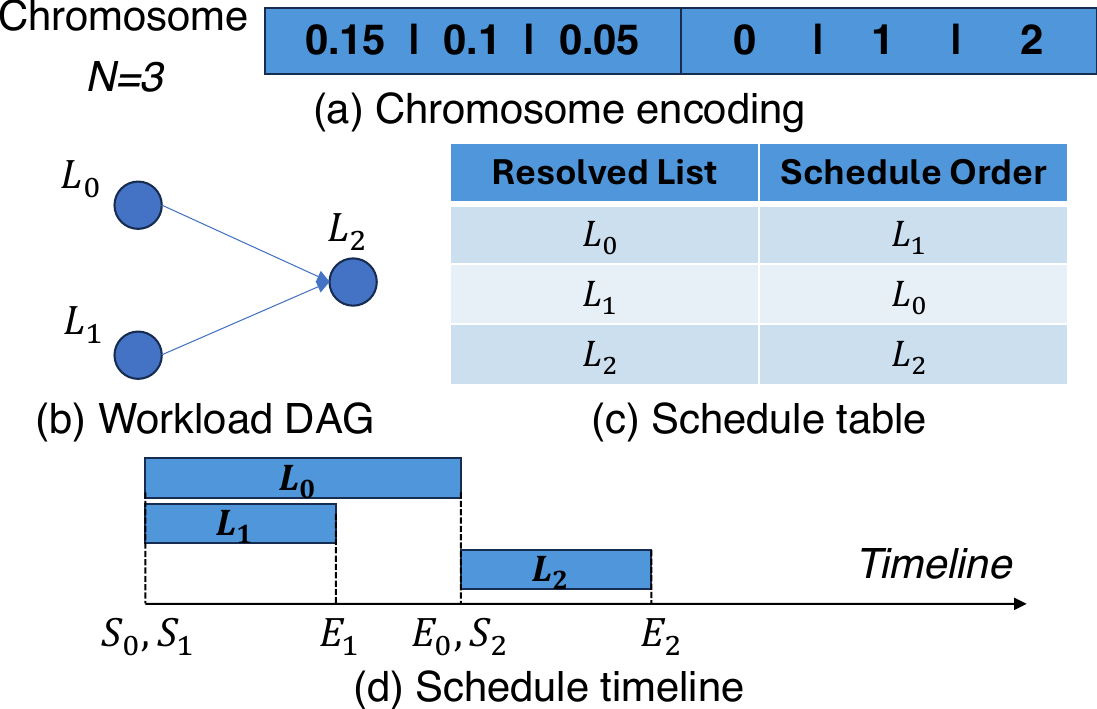}
    \vspace{-12pt}
    \caption{Illustration diagram for GA decoder.}
    \vspace{-12pt}
    \label{fig: GA decoder}
\end{figure}

Assume there is a child chromosome encoded as Figure \ref{fig: GA decoder}(a) and the workload DAG as Figure \ref{fig: GA decoder}(b). According to the data dependency, we can append $L_0$ and $L_1$ to the Resolved List in Figure \ref{fig: GA decoder}(c), then we search for the layer with a smaller chromosome Encode[$i$]. In this case, Encode[$1$] is smaller than Encode[$0$], thus, we append $L_1$ in the Schedule Order List. Then we iteratively check the dependency resolution and append the layer with the smaller Encode[$i$] to generate a complete Schedule Order List. Then, we start to schedule layers on the timeline following the order shown in Figure \ref{fig: GA decoder}(d), to explore the parallel execution under resource constraints. After generating the schedule timeline, we can get the makespan for the chromosome and evaluate its fitness, keep the chromosomes with the best fitness, and iterate for the next generation.

\vspace{-5pt}
\section{Experiment Results}

In this section, we first illustrate the single AIE efficiency in Section \ref{sec: single AIE experiment}, and design a series of MM workloads to show FILCO performance robustness on diverse workloads in Section \ref{sec: synthetic workload experiment}. We also conduct ablation experiments on the end-to-end throughput from BERT-32 to BERT-512 in Section \ref{sec: end2end experiment}. We evaluate our DSE algorithms in Section \ref{sec: DSE experiment}.
For baseline works, we apply the CHARM framework to obtain experiment results, and we build an in-house RSN analytical model for experiments, since RSN does not provide an analytical model.
All experiments are conducted on VCK190~\cite{versal_acap} with 150MHz on PL and 1GHz on AIE. AMD/Xilinx Vitis version 2023.1 is used as the compilation backend tool. For DSE, we leverage CPLEX and Pymoo to solve MILP and GA problems, respectively.

\vspace{-5pt}
\subsection{Single AIE Kernel Efficiency Comparison}
\label{sec: single AIE experiment}

\begin{figure}
    \centering
    \includegraphics[width=0.8\linewidth]{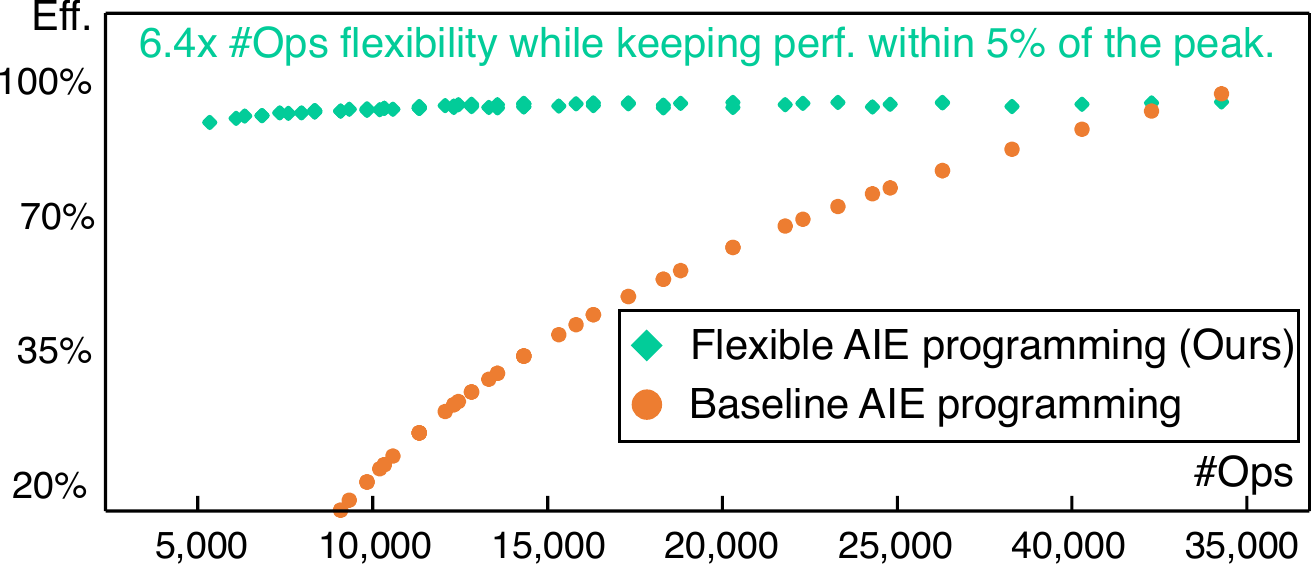}
    \vspace{-12pt}
    \caption{Single AIE efficiency under \#operations variation.}
    \label{fig: single_aie_eff}
    \vspace{-15pt}
\end{figure}

\jz{In this section, we illustrate the single-AIE computational efficiency gains of our proposed Flexible AIE programming using the FP32 MM benchmark of varying sizes.}
We apply the AIE intrinsics~\cite{amd_aie_intrinsics_ug} to program the single AIE kernel and measure the execution cycles on the Versal ACAP AI Engine System C simulator \cite{amd_aie_systemc_sim}.
As shown in Figure \ref{fig: single_aie_eff}, we evaluate the MM size across from 8x24x16 to 32x32x32 with the granularity of an atomic operation, i.e., 2x8x8. The results show that our design can sustain diverse MM sizes ranging from 14x24x16 to 32x32x32, achieving over $6\times$ variation in operation counts with only 5\% efficiency loss. In contrast, static AIE programming induces much more overhead due to data padding, causing a huge performance drop when the MM size is small.

\vspace{-5pt}
\subsection{Performance Comparison on Diverse MM}
\label{sec: synthetic workload experiment}

\begin{figure}
    \centering
    \includegraphics[width=0.75\linewidth]{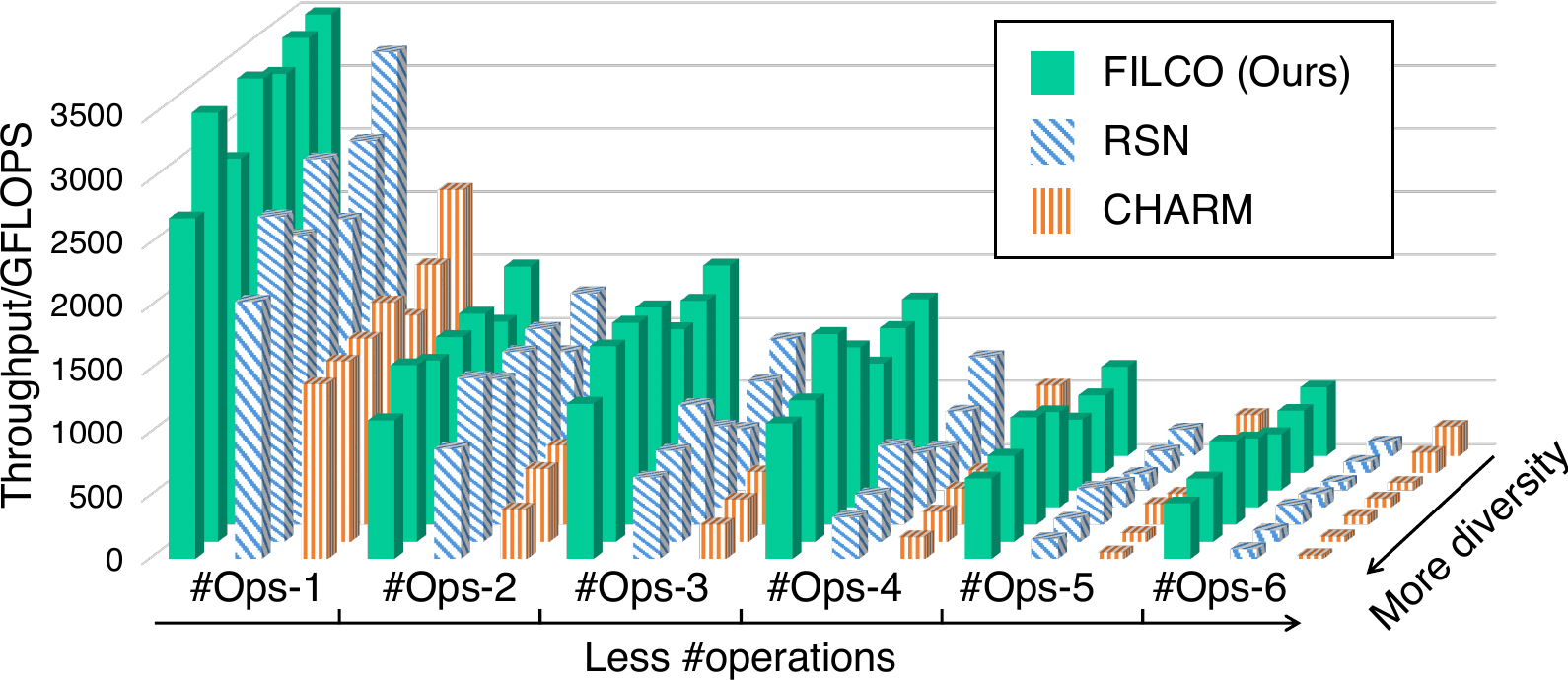}
    \vspace{-12pt}
    \caption{Throughput comparisons on diverse MM workloads.}
    \label{fig: MMexp}
    \vspace{-12pt}
\end{figure}

In this section, we design a series of Transformer-based workloads with \jz{varying} \textit{sequence length}, \textit{number of heads}, \textit{head dimension}, and \textit{MLP ratio}. Then, we categorize them according to the number of operations and inter-layer diversity, as shown in Figure \ref{fig: MMexp}.
When the operation count is large and the diversity degree is small, both CHARM and RSN can sustain a relatively high performance. \jz{As the diversity degree increases}, \jz{due to} the larger MM shape \jz{variance}, the performance drops sharply in CHARM, since it has to pad operand matrices to fit on-chip buffer shape, which incurs much more off-chip communication overhead. RSN can \jz{support} efficient resource utilization but only when the operation counts are large. This is because RSN can flexibly map operand matrices into different on-chip memory units only with fixed matrix shapes. As the MM size decreases, each memory unit incurs extra data padding, resulting in under-utilized computation resources and communication overheads for padded operands.
FILCO can sustain a large range of workload diversity since it can flexibly adjust the computation tile size and on-chip memory views to perfectly match operands and avoid resource under-utilization. In large MM workloads with a low degree of diversity, FILCO can achieve 1.3x throughput gains, and in small MM workloads with higher diversity, FILCO can achieve more than 5x throughput gains compared with RSN and CHARM.


\vspace{-5pt}
\subsection{Performance for Realistic Workloads}
\label{sec: end2end experiment}

\begin{figure}
    \centering
    \includegraphics[width=0.6\linewidth]{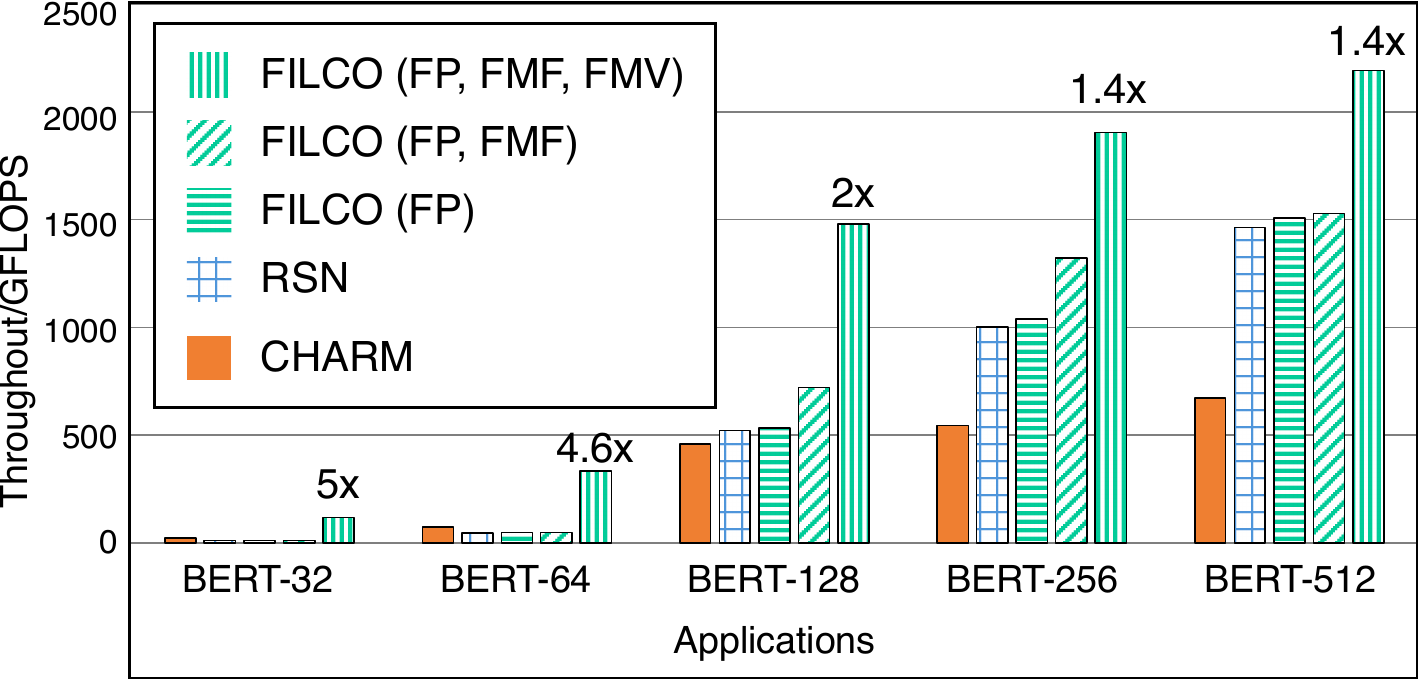}
    \vspace{-10pt}
    \caption{End-to-end performance of realistic BERT models. FP: flexible parallelism; FMF: flexible memory functionality; FMV: flexible on-chip memory views.}
    \label{fig: end2end}
    \vspace{-14pt}
\end{figure}

\begin{figure}
    \centering
    
    \includegraphics[width=0.6\linewidth]{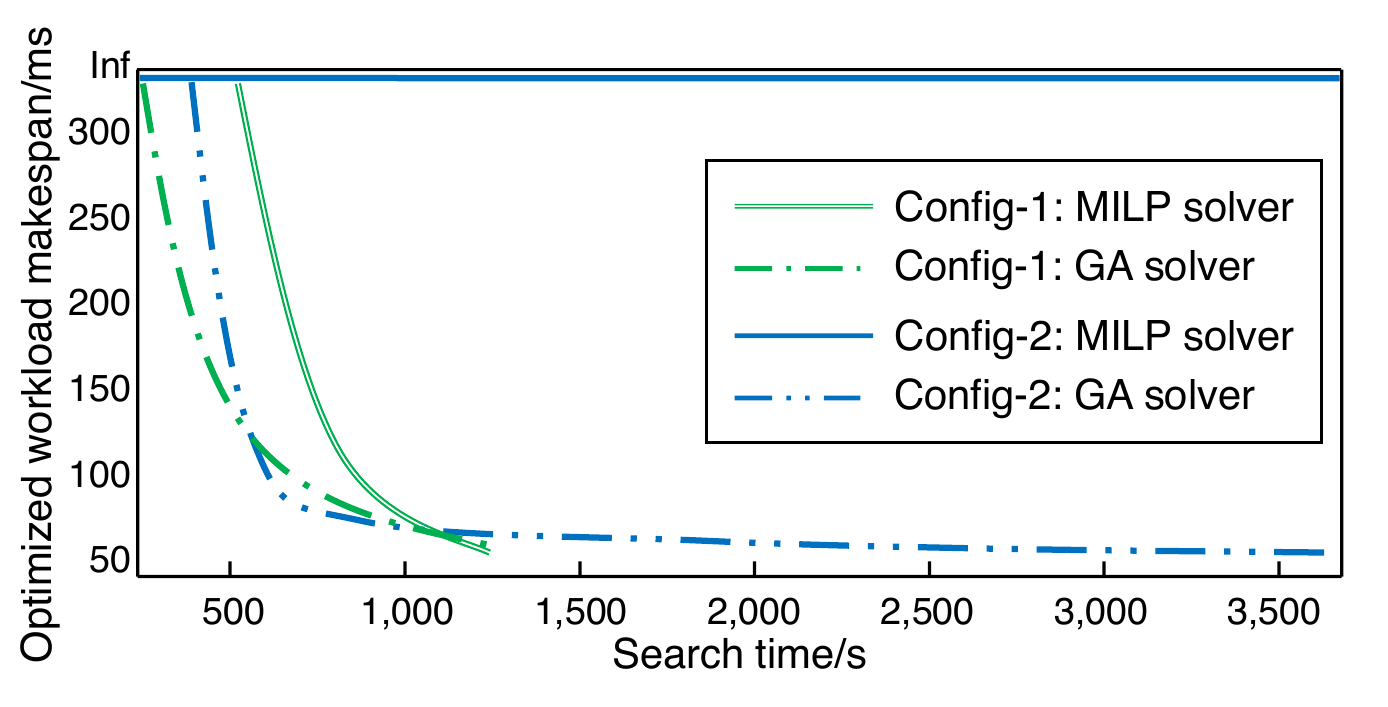}
    \vspace{-12pt}
    \caption{Comparison of search time for MILP and GA solver.}
    \label{fig: MILP GA compare}
    \vspace{-12pt}
    
\end{figure}

To evaluate the FILCO adaptability under diverse workloads and showcase the effectiveness of design methodology, we apply the FILCO framework to a series of BERT models: BERT-32, BERT-64, BERT-128, BERT-256, and BERT-512.
As shown in Figure \ref{fig: end2end},
we evaluate three FILCO designs with different enabled features.
For the small BERT applications, limited by a low CTC ratio, the communication time dominates the end-to-end throughput. Since RSN and CHARM are not able to adjust the tile size in one memory unit, data padding induces wasteful bandwidth utilization. FILCO (FP, FMF) and FILCO (FP) are also inefficient, as they load many padded operand matrices, incurring significant communication overhead. While FILCO (FP, FMF, FMV) can efficiently handle it since FMUs can flexibly adjust the buffer view to perfectly match the small matrices without data padding.
In the large BERT applications, CHARM and RSN can achieve a relatively high efficiency. However, since RSN does not explore the design space for diverse parallelism patterns, and the functionalities of on-chip memory are fixed before compilation time, this leads to a sub-optimal design point. After enabling these features, FILCO can achieve better design points compared with RSN and CHARM.

\vspace{-5pt}
\subsection{Evaluation for DSE Search Efficiency}
\label{sec: DSE experiment}

We evaluate the search time to demonstrate the \jz{efficiency and scalability of our proposed MILP and GA-based DSE methods.}
As shown in Figure~\ref{fig: MILP GA compare}, we compare the MILP and GA search time under two task sets. \textit{Config-1} represents a workload with 50 layers, each with 50 candidates, whereas \textit{Config-2} is a workload with 50 layers, each with 5000 candidates.
In the small task set, the GA algorithm achieves a near-optimal solution with a faster convergence rate than MILP, with only about a 3\% optimality gap. In the large task set, GA is able to produce a good design point within 10 minutes, whereas MILP fails to obtain a valid solution even after one hour. \jz{Therefore, for small workloads, FILCO can obtain optimal schedules using the proposed MILP formulation. For larger workloads, FILCO GA delivers near-optimal solutions with significantly shorter search time, showing strong efficiency and scalability.}


\vspace{-10pt}
\section{Related Work}

Existing works have two trends to explore hardware efficiency for diverse workloads: fixed-dataflow accelerators~\cite{zhuang2023charm,ssr,chen2024understanding,hall2020hpipe,liu2025flightvgm,zeng2024flightllm,dong2024eq,zhang2020dnnexplorer,kwon2021herald,cai2023set,iccad23aim,glsvlsi25art,rtss25clare} and overlay-based accelerators~\cite{RSN,tong2024feather,yang2025nsflow,he2025intar,abdelfattah2018dla,zhang2022fast}.
Fixed-dataflow accelerators are efficient in static and lower-diversity scenarios, since they must explicitly consider the datapath for every possible execution pattern, inducing much overhead and resource under-utilization. CHARM~\cite{zhuang2023charm} designs a two-diverse accelerator, but they still require extra data padding when the hardware does not match workloads. 
DNNExplorer~\cite{zhang2020dnnexplorer}  designs a pipeline-generic hybrid architecture, which also fixes the datapath and does not alleviate the problems.
SSR~\cite{ssr} and EQ-ViT~\cite{dong2024eq} cannot handle applications with large model sizes, limited by their execution patterns.
FlightVGM and FlightLLM~\cite{liu2025flightvgm,zeng2024flightllm} suffer from performance degradation on applications except for LLM. Herald~\cite{kwon2021herald} provides a coarse-grained flexibility, and spatially implementing accelerators induces performance degradation when batch numbers are small.

Overlay-based accelerators design a flexible datapath and apply token-based control to enable different datapaths for diverse workloads. 
InTAR~\cite{he2025intar} designs a runtime fixed datapath, and the functionalities of each PE cannot be reconfigured.
FEATHER~\cite{tong2024feather} mainly focuses on data layout reordering.
NSFlow~\cite{yang2025nsflow} is limited by on-chip memory management, as the dataflow between PEs and on-chip memory is static at runtime. 
RSN~\cite{RSN} alleviates the problem to some extent, but their computation parallelism is static, and on-chip memory management is not flexible enough, incurring much performance degradation due to computation under-utilization and communication overhead.

Memory management is key to improving hardware utilization.
GraDMM~\cite{wang2024scalable} proposes a dynamic memory management library for HLS-based FPGAs.
Beyond on-chip techniques, several works~\cite{wong2023dongle, qureshi2023gpu, chang2024gmt, yang2025agile} extend accelerator memory hierarchies to incorporate off-chip DRAM and storage systems to address the rapid growth of model sizes. Other works \cite{aries,hunhoff2025iron,chen2024allo} also explore the programming abstraction for productivity.
We consider these directions complementary to FILCO and leave them as future work.
\vspace{-10pt}
\section{Conclusion}

In this paper, we propose the FILCO architecture and framework to provide a flexible composing architecture that can effectively improve hardware efficiency on hugely diverse workloads. We will keep on exploring FILCO in more diverse scenarios in future works.
\smallskip
{\small
{\noindent\textbf{ACKNOWLEDGEMENTS --}} This work is supported in part by Brown University New Faculty Start-up Grant, NSF awards 
\#2140346, 
\#2231523, 
\#2441179, 
\#2348306, 
\#2511445, 
\#2518375, 
\#2536952. 
We thank AMD for the hardware and software donations.

\bibliographystyle{ACM-Reference-Format}

\bibliography{Ref_X}

\end{document}